\documentclass[letterpaper,twocolumn,10pt]{article}
\usepackage{usenix,epsfig,endnotes}
\usepackage{graphicx}
\usepackage{algorithm}
\usepackage{algorithmic}

\begin{document}

\title{\Large \bf A Symphony Conducted by BruNet}
\author{{\rm P. Oscar Boykin}\\
Department of Electrical and Computer Engineering\\
University of Florida\\
\and
{\rm Jesse S. A. Bridgewater}\\
Electrical Engineering Department\\
University of California, Los Angeles\\
\and
{\rm Joseph S. Kong}\\
Electrical Engineering Department\\
University of California, Los Angeles\\
\and
{\rm Kamen M. Lozev}\\
Electrical Engineering Department\\
University of California, Los Angeles\\
\and
{\rm Behnam A. Rezaei}\\
Electrical Engineering Department\\
University of California, Los Angeles\\
\and
{\rm Vwani P. Roychowdhury}\\
Electrical Engineering Department\\
University of California, Los Angeles\\}

\date{}

\maketitle

\begin{abstract}
We introduce BruNet, a general P2P software framework which we use to produce
the first implementation of Symphony,
a 1-D Kleinberg small-world architecture.
Our framework is designed to easily implement and measure different
P2P protocols over different transport layers such as TCP or UDP.
This paper discusses our implementation of the Symphony network,
which allows each node to keep $k \le \log N$
shortcut connections and to route to any other node with a short
average delay of $O(\frac{1}{k}\log^2 N)$.
We present experimental results 
taken from several PlanetLab deployments of size up to 1060 nodes.  These successful 
deployments represent some of the largest PlanetLab deployments of P2P overlays found 
in the literature, and show our implementation's robustness to massive node
dynamics in a WAN environment.
\end{abstract}

\section{Introduction: Motivation and Summary of Results}\label{sec:intro}
Peer-To-Peer (P2P) networking is an increasingly popular network model where nodes
communicate directly without utilizing a centralized server.
In recent years, P2P file-sharing applications have flourished.
A recent study shows 
that P2P systems are responsible for approximately one half of the network traffic at a 
major university\cite{ss:P2Pmeasurement} and comprise a significant fraction
of total Internet traffic.  For a review of P2P search systems, see \cite{rm04}.

There are three novel contributions reported in this work.  First, we describe
a new P2P software framework called BruNet.  The BruNet framework handles most
of the issues common to all P2P protocols such as dealing with firewalls and
NATs, connecting nodes, and routing packets.  Secondly, we use the BruNet P2P
framework to implement Symphony\cite{pr:Symphony}, a 1-D Kleinberg routable
small-world network\cite{jk:SmallWorld, jk:Algorithmic}.  This is the first
implementation of a 1-D routable small-world network.  Third, we report on
large scale PlanetLab tests involving more than 1000 nodes, which puts the P2P
networks described here amongst the largest P2P networks to be tested on
PlanetLab. These deployments demonstrate our implementation's robustness to
massive node dynamics in a WAN environment.

Our BruNet software architecture manages P2P packet routing and
connection maintenance.  Given a
packet with a particular
destination address $A$, the system will deliver the packet to the node closest
to that address.  This sort of routing primitive may be used to build a
distributed hash table (DHT), which is common in the P2P literature.
Clearly, the success and efficacy of such an ad-hoc addressing and routing scheme
depends on the robustness of the overlay structured networks.

The deployment of DHT P2P systems such as the Kademlia-based\cite{pm:Kademlia}
eDonkey, which already supports about a million simultaneous users, indicates
that large-scale overlay networks are feasible.  The existence of such
large-scale DHT systems is impressive, however the performance of P2P networks
at that scale has not yet been systematically studied.  While we have not
yet scaled to one million nodes, our experiments of more than 1,000 nodes is
amongst largest P2P networks to be tested on PlanetLab.  The data we
obtained from deployments of our system on PlanetLab show that the structured
routing network can indeed be bootstrapped from a random initial network, and
can be robust to high rates of joins and departures of participating nodes.

We chose Symphony, the 1-D Kleinberg routable 
small-world network\cite{pr:Symphony, jk:SmallWorld, jk:Algorithmic} as the topology for
the structured overlay network. This ringlike address space entails simple
routing calculations and requires very low node state.
Our structured overlay is currently the only implementation of a 1-D
Kleinberg routable small-world network; as reviewed in the next section, a
number of schemes that utilize the 1-D small-world model have been proposed,
but to the best knowledge of the authors none have been deployed and tested in 
a WAN environment. Kleinberg proved that properly designed small-world
networks  could support efficient decentralized routing with
$O(log^2 N)$ latency. The proposed system uses a 160-bit address space to
construct a ringlike structure.  Shortcuts are made in this ringlike address space
according to a specific probability distribution\cite{jk:SmallWorld}.
The analysis and simulation
results in \cite{pr:Symphony} show that maintaining $k \le \log N$
long-range neighbors improves routing latency to $O(\frac{1}{k} \log^2 N )$.

Our functioning implementation adds several new features to the routable
small-world model, including expanded routing rules to permit firewall traversal and easy
bootstrapping and also to obtain a structured 1-D ring starting from any
initially connected network. Networks  up to 1060 nodes have been deployed on PlanetLab,
as we discuss later in Section \ref{sec:experimental_results}. A key goal of
this effort is that the network remains routable in the presence of massive
node dynamics including massive joins, massive failures, ring merging and
churn. The system's robustness under heavy node dynamics compares very favorably to
the results published for Tapestry\cite{bz:Tapestry}; moreover, our deployment
has more than twice the number of nodes dealt with in \cite{bz:Tapestry}.

The paper is laid out as follows: we first discuss related work in the following
section.  Section \ref{sec:overview} describes
the BruNet software architecture and system components.  Section
\ref{sec:overview} also includes our approach to traversing firewalls and
NAT devices.
Section \ref{sec:structured_arch} provides details on our Symphony
implementation.  Finally, Section
\ref{sec:experimental_results} presents PlanetLab experiments that
demonstrate the correctness and robustness of the network.

\section{Related Work}\label{sec:related_work}
There has been much recent work on producing structured P2P overlays
with distributed hash table (DHT) interfaces. Some examples of these
structured systems include
\cite{stoica-2001,is:Chord,sr:CAN,ar:Pastry,dm:Viceroy,nh:SkipNet,
bz:Tapestry,sr:Bamboo,pm:Kademlia,pr:Symphony}.
The main advantages of these structured DHT systems are scalable object location in
 $O(\log N)$ or $O(\log^2 N)$ steps and the guaranteed retrieval of any existing object.

This paper reports on an implementation and measurements
rather than simulation of a P2P network.
While there are many reports of simulations of structured P2P protocols, the measurement
of such protocols in real world WAN environments has rarely been
addressed (e.g. \cite{bz:Tapestry}).

Among the existing structured systems, there are several Kleinberg-inspired
small-world P2P overlays: Symphony \cite{pr:Symphony} provides a detailed
software design for a DHT system based on a unit-circumference ring; Accordion
\cite{jl:Accordion} is a proposed small-world-based structured system designed
to provide efficient bandwidth management of the distributed routing tables;
Mercury \cite{ab:Mercury} presents a protocol for supporting multi-attribute
range queries that layers on top of a small-world-based ring; SWAN
\cite{eb:Swan} is an implemented multi-agent system based on the original 2-D
Kleinberg model \cite{jk:Algorithmic}.  Of the aforementioned small-world P2P
systems, only SWAN has been implemented, while performance estimates for
Symphony, Accordion, and Mercury are based solely on simulations.  Therefore
the presented system appears to be the first implementation of the 1-D
ring-based Kleinberg routable small-world network.

\section{BruNet System Architecture}
\label{sec:overview}
The BruNet P2P software framework is designed to allow easy implementations of many
different protocols.  The software is implemented in the $C\#$ programming
language using the Mono compiler and virtual machine on GNU/Linux based systems.
This section provides a general overview of the basic primitives of the system,
namely nodes, addresses, edges, routers and connection overlords.

\begin{figure}
\centering
\includegraphics[scale=0.4]{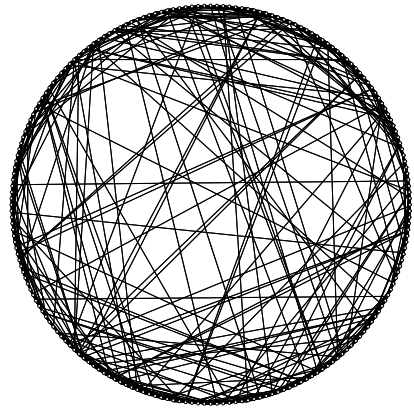}
\caption{The structured ring permits efficient routing
between nodes. This 200-node network was run on PlanetLab.}
\label{fig:network_structure}
\end{figure}

\subsection{Nodes and Addressing}
\label{sec:BruNet_nodes}
The active elements in the system are called nodes.  Each node can send
packets, receive packets, and route packets.  A particular computer system,
such as a desktop PC or a server system may host one or more
nodes.  The node is envisioned as an agent for a user or software application.
Each node has exactly one address, which uniquely identifies that node
on the network.  Additionally, each node maintains several edges and uses
these edges to pass packets to neighboring nodes.

When a node is the destination of a packet, the node informs the user, or a
higher-layer software application, of the packet.  The node also acts as a
manager of its edges.

The 160-bit address space consists of all the integers from $0$ to
$2^{160}-1$ and is partitioned into 161 distinct address classes.
To determine the class of a particular address, count the number
of consecutive bits of value $1$ on the rightmost part of the address.  There can be
between 0 and 160, and thus there are 161 address classes.  Clearly,
address class $n$ is twice as large as $n+1$.
In fact, a class $n$ address ends with
exactly one bit of value $0$ followed by $n$ bits of value $1$ (except for
class 160, for which
all bits have the value $1$).  The size of the class $n$ address space is $2^{159-n}$
(except class 160, which has size 1).  To see that we have accounted for all the
addresses, we can sum the size of each class and see that we get all $2^{160}$
addresses:
\begin{eqnarray*}
S &=& 1 + \sum_{k=0}^{159} 2^{159 - k}\\
  &=& 1 + 2^{159}(\frac{1 - 2^{-160}}{1 - 2^{-1}})\\
  &=& 1 + 2^{160} - 1 = 2^{160}
\end{eqnarray*}
So we see that if we count all classes from $0$ to $159$ (and add $1$ for
class 160), we see that we get all $2^{160}$ possible addresses.

Address class-$0$ is the largest.
We use class-$0$ to represent addresses on the ring.  These ``ring" addresses are common
to both the Chord\cite{is:Chord} and Symphony\cite{pr:Symphony} protocols.
We describe the routing algorithm for these addresses in
Section \ref{sec:structured_arch}.
In addition to the ring addresses in class-$0$, we define
class-$124$ as ``directional" addresses.  Directional addresses indicate that a
packet should be routed in a particular direction on the ring
such as clockwise or counter-clockwise.  Directional addresses are useful
for communicating with nearby nodes on the ring as is often needed when
joining the network or in DHT applications.

Our system is designed to be a general framework for P2P applications.
For example, one application of our system might be to use
class-$1$ addresses to represent
hypercube addresses such as those used in the Pastry P2P protocol
\cite{ar:Pastry}.  This partitioning allows us to easily implement
new protocols without changing the packet format or core libraries.

\subsection{Packet Format}
\label{sec:BruNet_packets}
All system packets begin with a byte that describes
the type of data contained in the payload, followed by a payload.
The first packet type is $0x01$, which is used by nodes
to establish connections and discover one another's BruNet system
information.

The second packet type is $0x02$, which is used
for the routed P2P protocols (this type is in contrast to type $0x01$ packets which
are not routed on the overlay and are only used when two nodes are
directly connecting to one another).
In many respects, the routed P2P packets are similar to Ethernet
packets but with a few notable differences.
Ethernet has 8 byte addresses where this system uses 20 byte addresses.
Ethernet uses two bytes to denote the payload type, where we use only
one.
Unlike Ethernet packets, we
do not need to include a checksum (since, as we discuss in section
\ref{sec:BruNet_packet_transport}, we assume that the edges provide accurate
packets).  Also unlike Ethernet, we \emph{do} need to include a field to
indicate how far the packet has traveled and how far it is allowed to go.

\begin{table}
\begin{tabular}{l|l|l}
Header Field & Start Position& Length (bytes)\\
\hline\\
Type & 0 & 1\\
Hops & 1 & 2\\
TTL & 3 & 2 \\
Source & 5 & 20\\ 
Destination & 25 & 20\\ 
Payload Type & 45 & 1 
\end{tabular}
\caption{Packet format}
\label{tab:ahpacket}
\end{table}

Packets may encapsulate many different types of payloads.
For instance, nodes manage their position in the network by sending
``network structure" packets to other nodes.
Packets also transport what may be considered ``application layer" data, such
as queries for DHT or file-sharing applications.

\subsection{Edges and Connectivity}
\label{sec:BruNet_packet_transport}
In this work, we will say that a pair of nodes
has an edge between them if they are communicating with one another by
sending packets over a single overlay hop.
Any underlying
networking protocol which matches this requirement is a suitable transport.
In fact, different edges may work over different transport protocols
(such as TCP, UDP, etc.).

Every edge must provide two things:
\begin{itemize}
\item the edge must not pass corrupt packets
\item the edge must know the length of each packet it receives.
\end{itemize}

We identify endpoints of edges with transport addresses, for instance
\emph{BruNet.tcp:192.168.0.1:10030} to identify an endpoint of a TCP
edge at IP address 192.168.0.1 and port 10030.  Generally, the transport
address is a pair which contains the protocol and the addressing information
for that protocol.  Currently, we have implemented TCP and UDP edges, but
in principle we could also define an Ethernet edge to transport BruNet
over Ethernet.

Edges are typed with labels.  For instance, in the Symphony protocol, there
are edges which go to near neighbors on the ring and also shortcut connections
that cut across the ring.  The edges are labeled to distinguish them.  Our
framework allows edges to be labeled with any string, so a future protocol may
be implemented which may define new edge labels.

We assume that each node joins the network by contacting some node and forming
a ``leaf" connection.  The leaf connection is used for a newly joined node to bootstrap
into its proper place in the network.  The new node bootstraps by asking the node
on the other end of the leaf connection to act as a proxy for any packets the
new node would like to send or receive.  Once a node has at least one leaf
connection, it may use that connection to get more connections.  There are two
connection phases: making a connection request and the handshaking which
goes on when two nodes are creating an edge between them.

Consider the case of one node, which we will call the source, connecting with
a second, which we will call the target.
To create a new connection, the source sends a message to the target \emph{through
the BruNet network}.  This message includes the BruNet address as well as
a list of transport addresses corresponding to the source node.
Once the target node
receives the connection request, it sends a response which includes the same
information about the target, namely the target's BruNet address and list
of transport addresses.
After sending the response, the target also
attempts to create a new edge by using some networking transport to contact
the source node.  For instance, when the source node is using UDP, the target
node will send a UDP packet to the address given in the connection request.
The target attempts to connect to the source using each item in the
transport address list.  If none of these attempts is
successful the target gives up.  On the other end of this exchange, the
source node should receive both a response to its connection request and
the new edge connection from the target.  Assuming the transport layer
is faster than the BruNet layer (which should be true since BruNet is
an overlay on the transport), the source node should get the target's
connection prior to receiving the response to the connection request.
If for any reason (such as the existence of a firewall which we discuss
in Section \ref{sec:firewalls}) the source does not get a connection
from the target, when it receives the target's connection message response,
the source initiates a connection to the target.

Assuming one or the other of the nodes is able to make a connection to the
other, the nodes connect and exchange several pieces of information, which
we call the linking protocol.
The first
piece of information the nodes exchange is the local and remote transport
addresses that each see as accurate for the connection.  Due to network
address translation (NAT), the two nodes may not agree on which IP addresses
and port numbers they are each using, but the information is exchanged
so that each node can add this new transport address to their list of possible
transport address endpoints that future nodes may use to connect to them.
In addition to two peers' transport address information, each node
exchanges a list of BruNet addresses (which are used for routing
on the overlay) and transport addresses (which are used for making new
connections) of nearby nodes.
In our experience, getting connected, sending and
receiving packets, and dealing with the errors that may occur during this
process is the most complex aspect of the P2P system.  As such it is very
convenient to design this aspect of the system to be reusable by a wide
variety of protocols.

\subsection{Firewalls}
\label{sec:firewalls}
Many nodes on the Internet today are behind a firewall or a network
address translation (NAT) device.  Such nodes present a challenge to P2P
systems as it can be difficult for them to become connected to the network and
to each other.
As we discussed in Section \ref{sec:BruNet_packet_transport},
the BruNet connection process involves two steps: sending the connection request
followed by the linking protocol.

When at least one node is not behind a NAT or a firewall, our standard
connection protocol will result in the nodes forming a connection between
them.
Since our connection protocol involves first contacting
the target over the BruNet network to exchange transport address information,
\emph{both} the target and the source have enough information to contact the
other.  So as long as one of the two parties is not behind a firewall, the
connection will take place normally.

When using UDP, our protocol allows two NATed and firewalled nodes to
connect.
As identified by the STUN\cite{STUN} protocol, there are four types of NAT
in use today: full cone, restricted cone, port restricted cone and
symmetric.  Like the STUN protocol, we only deal with the first three cases,
and not with the symmetric NAT.  Of the first three cases, the port restricted
cone is the most restrictive; any protocol that works for the port restricted
case works for the first two, so we describe how we deal with the port
restricted cone NAT.

A port restricted cone NAT performs a mapping from an internal network
$(IP_i,port_i)$ pair to an external $(IP_e,port_e)$ pair.
Consider a packet that arrives at the NAT with destination $(IP_e,port_e)$ and
source $(IP_s, port_s)$.  The NAT will only pass this packet if the internal
node $IP_i$ has previously sent a packet with source $(IP_i, port_i)$ to
$(IP_s, port_s)$.  So, in order for two nodes which are both behind
a NAT to communicate, \emph{both} nodes have to have previously sent a packet
to the other's translated address.  Fortunately, since our connection protocol
involves routing the transport address information over the overlay, both
nodes will get transport address information sufficient to contact the other.
Assuming that both know their translated addresses, each will be send packets
to the other's translated addresses.  If the NATs are not symmetric, they
will pass all packets after the first.  Our linking protocol
involves using retries with back-off, thus the nodes will be able to send
the necessary packets to open the connection through the NAT.  The only issue
that remains is how nodes learn their translated transport address.
As covered in Section \ref{sec:BruNet_packet_transport}, part
of our protocol is for each node to echo the transport address information it
sees to its peer during connection.  This allows each node to learn its
translated address assuming it can make at least one leaf connection to a node
which is not behind a NAT.

Our approach uses the same facts about common NAT devices as the STUN protocol
except we use the P2P network instead of a
central server to share the translated IP information.

\subsection{Routing and Connection Management}
Most P2P systems will have a great deal of overlap in the
concepts we have discussed above, however significant differences
will emerge when it comes to routing of packets and the management of
connections to peers.  In the BruNet architecture, both routing and connection
management are handled by components.

To implement a new protocol, most of the existing BruNet system is reused, but
a new router object must be defined and associated with the address class that will
be used for that protocol.  Additionally, each P2P protocol may have different
rules for maintaining connections to peers including how many connections to
maintain and to which peer each node should be connected.  Connection overlord 
objects encapsulate the code which manages the connections in the system.  For
instance, in the Symphony protocol, each node should have a connection to
its left and right neighbors as well as at least one shortcut connection.
We implemented a \emph{SymphonyConnectionOverlord} which counts the number
of each of these types of connections, initiates new connections when needed,
and closes connections that are no longer needed.

BruNet was designed to implement unstructured as well as
structured P2P protocols.  Implementing unstructured protocols, such as
the Gnutella broadcast query protocol, is also easy.  One need only define
a new address class to represent broadcasts, implement a router to handle
the routing of the broadcast messages and to build a routing table of known
addresses, and finally a connection overlord that makes sure that the node
stays connected to the network as nodes come and go.  The connection logic,
transport abstraction, packetization, and serialization can all be reused between
various implementations.

\section{An Implementation of Symphony}
\label{sec:structured_arch}
In the previous section we discussed the architecture of the BruNet P2P
framework.
In this section we describe our implementation of the Symphony 1-D small-world
system.
To implement a particular P2P system, we need to describe
the routing and connection management, including joining and leaving,
which we discuss in Sections \ref{sec:structured_routing} and
\ref{sec:structured_joining}  respectively.

We use class-$0$ addresses for this protocol.  Thus, each node in the
network can take one of $2^{159}$ structured
addresses\footnote{for randomly selected  addresses, the network size will have
to be $\approx 2^{79}$ nodes before we are likely to reuse an address}.
We interpret these addresses as even 160-bit integers in the range
$[0,2^{160}-2]$ with this address space forming a ring.  By convention, we say that the
ring \emph{increases} in the \emph{clockwise} direction.

\subsection{Small World Routing}
\label{sec:structured_routing}

The theory that supports structured routing comes from works on
routable small-worlds \cite{jk:SmallWorld,jk:Algorithmic}. However, we introduce novel practical
routing algorithms, which make network maintenance a natural consequence of those
routing algorithms.  As we discuss in Section \ref{sec:BruNet_nodes}, each
node has an address that can be interpreted as a coordinate on a ring.  As
such, there is directionality (e.g. clockwise and counterclockwise).  There
are two mechanisms for routing on this structure: destination based and
direction based.

\begin{algorithm}
\begin{algorithmic}
\caption{$GreedyNextHop(v,source,target)$: This algorithm describes how a packet
arriving at $v$ from {\bf $source$} takes its next hop towards the {\bf
$target$}  using greedy mode.
Each hop tries to get closer (without visiting {\bf $source$} ) to {\bf $target$}.
The adjacency list of node $v$ is denoted Adj[v], and the
distance between two nodes (a,b) in the network is
$DIST_{ring}(a,b)$.}

\label{alg:struct_greedy}
  \STATE $d_{min} \Leftarrow DIST_{ring}(v,target)$
  \STATE $u_{min} \Leftarrow v$
  \FORALL{$u \in Adj[v]$}
  \STATE $d_{tmp} = DIST_{ring}(u,target) $
  \IF{$d_{tmp} < d_{min}$}
  \STATE $d_{min} = d_{tmp}$
  \STATE $u_{min} = u$
  \ENDIF
  \ENDFOR
  \IF{$u_{min} \neq v$ or $u_{min} \neq source $}
  \STATE Deliver to $u_{min}$
  \ELSE
  \STATE This is the last hop. Deliver locally to $v$.
  \ENDIF
\end{algorithmic}
\end{algorithm}

\begin{algorithm}
\begin{algorithmic}
\caption{$ExactNextHop(v,source,target)$: This algorithm describes how a packet
arriving at $v$ from {\bf $source$} takes its next hop towards the {\bf
$target$}  using exact mode.
Each hop tries to get closer (without visiting {\bf $source$} ) to {\bf $target$}.
The packet is delivered only to the target and no other node.
The adjacency list of node $v$ is denoted Adj[v], and the
distance between two nodes (a,b) in the network is
$DIST_{ring}(a,b)$.}

\label{alg:struct_exact}
  \STATE $d_{min} \Leftarrow DIST_{ring}(v,target)$
  \STATE $u_{min} \Leftarrow v$
  \IF{$v == target$}
  \STATE This is the last hop. Deliver locally to $v$.
  \ELSE
  \FORALL{$u \in Adj[v]$}
  \STATE $d_{tmp} = DIST_{ring}(u,target) $
  \IF{$d_{tmp} < d_{min}$}
  \STATE $d_{min} = d_{tmp}$
  \STATE $u_{min} = u$
  \ENDIF
  \ENDFOR
  \IF{$u_{min} \neq v$ or $u_{min} \neq source $}
  \STATE Deliver to $u_{min}$
  \ENDIF
  \ENDIF
\end{algorithmic}
\end{algorithm}

In direction based routing, we use fixed addresses (class-$124$) to refer to 
``clockwise" and ``counterclockwise".  When the packet's HOPS equal its TTL, 
the packet is delivered.  By setting the TTL, a node can then communicate with 
its near-neighbors on the ring.  This might have interesting applications for
caching in DHT systems.  Nodes maintain connections to at least two nearest
nodes to them in both directions. This direction based routing is what enables a node
to find its near-neighbors in order to connect to them.

Destination based routing is slightly more complex.  This mode of routing
refers to the case where one node wants to address a second node by that
second node's class-$0$ address, not based on its relative position on the ring.  
The simplest approach would be to route to the neighbor node which is closest to 
the destination, never routing to a node that is further.  This routing type
is described in Algorithm \ref{alg:struct_greedy}. Clearly there can be
no loops since each packet must get closer to the destination at each step. In
some cases it may be desirable for a packet to only be delivered to the exact
target class-$0$ address as shown in Algorithm \ref{alg:struct_exact}.
Kleinberg showed that the number of hops is $O(\log^2 N)$ on average between any
two nodes (when each node has $1$ correctly distributed ``shortcut"
connection)\cite{jk:Algorithmic,jk:SmallWorld,jk:Information}. If $k \le \log N$ ``shortcuts'' are maintained, the routing
latency can be reduced to $O(\frac{1}{k}\log^2 N)$ hops. This result allows
for a trade-off between node degree and routing latency.

\begin{algorithm}
\begin{algorithmic}
\caption{$AnnealingNextHop(v,source,target)$: This algorithm describes how a packet
arriving at $v$ from {\bf $source$} takes its next hop towards the {\bf $target$}  using annealing mode.
Each hop tries to get closer (without visiting {\bf $source$} ) to {\bf $target$} unless that is not possible
in which case the packet is delivered to $v$ and sent to the next closest node. The
adjacency list of node $v$ is denoted Adj[v], and the
distance between two nodes (a,b) in the network is
$DIST_{ring}(a,b)$.}
\label{alg:struct_annealing}
  \STATE $d_{min} \Leftarrow DIST_{ring}(v,target)$
  \STATE $d_{sec} \Leftarrow d_{min}$
  \STATE $u_{min} \Leftarrow v$
  \STATE $u_{sec} \Leftarrow v$
  \FORALL{$u \in Adj[v]$}
  \STATE $d_{tmp} = DIST_{ring}(u,target) $
  \IF{$d_{min} \le d_{tmp} < d_{sec}$}
  \STATE $d_{sec} = d_{tmp}$
  \STATE $u_{sec} = u$
  \ELSIF{$d_{tmp} < d_{min}$ }
  \STATE $d_{sec} = d_{min}$
  \STATE $u_{sec} = u_{min}$
  \STATE $d_{min} = d_{tmp}$
  \STATE $u_{min} = u$
  \ENDIF
  \ENDFOR
  \IF{$u_{min} \neq v$ or $u_{min} \neq source $}
  \STATE Deliver to $u_{min}$
  \ELSE
  \STATE Deliver locally to $v$.
  \STATE Deliver to $u_{sec}$
  \ENDIF
\end{algorithmic}
\end{algorithm}

In a real system there may be some problems to deal with.  In particular, the
ring may be broken by several nodes leaving at once.  In that case, the ring
becomes a line.  If the line is not reconnected into a ring, a subsequent
failure could cause the line to split, which would break routability.  As
such, we add some exceptions to the simple routing discussed above which makes
reconnecting the ring easier: namely, we do not require that the packet gets
closer to its destination on its first hop as described in Algorithm
\ref{alg:struct_annealing}.

\subsection{Joining the Small-World}
\label{sec:structured_joining}
In order for a node to join the ring, it makes use of
Routing Algorithm \ref{alg:struct_annealing}:
annealing routing.  The annealing routing tolerates some disorder in the
network.  
Every node that joins
the ring must have a 160-bit class-$0$ address. This address
must be randomly-generated to ensure the near uniform distribution of addresses on the
ring; thus class-$0$ addresses are obtained by using a secure hash algorithm or 
some other source of random bits. After a node has a class-$0$ address, it must
find its place in the ring. This means that it needs to make a connection to
the closest node on both the right and left of its own address. Since the new
node is not yet connected to the correct place in the ring it is not yet able to
route messages using the routing algorithms described above.
The new node instead makes use of a node that is correctly placed in the ring as a
proxy in order to find its place. The new node creates a special type 
of bootstrapping connection that does not support any of the routing algorithms 
above but does provide for packets to be sent to the node on the other end of 
the connection. 
This bootstrapping connection allows the new node to communicate with the
proxy in order to send and receive messages while it is waiting to find its
place in the ring.
The proxy sends a request to connect to the new address which is not 
yet in the network.  Given the new node's absence, the closest node on the
right and the closest node on left of the new node will form connections to the
new joining node. At this point the new node is at the correct location in the
ring and can add additional neighbors and shortcut connections as needed.
Algorithm \ref{alg:struct_join} shows this process.

Connection is not an instantaneous process.  Our implementation uses two round
trips: a link request and response, and a status request and response.  The
link messages exchange the node addresses, the IP addresses and port
numbers, and whether the connection is a near-neighbor connection or shortcut
connection.  The status message allows the nodes to communicate some of their
properties to their neighbors.  In particular, the status message shares the
node address and IP information of other nodes which are close to the new
neighbor.  This information allows nodes to verify that their views of the
network are consistent and make repairs.

\begin{algorithm}
\begin{algorithmic}
\caption{$JoiningTheRing(v,u)$: This algorithm describes how a new node,
denoted as $v$, joins the structured ring. The proxy that helps $v$ find its
place in the network is called $u$. The class-$0$ address of a node is
denoted as $ADD(node)$. $ADD(v_c)$ is the closest address to $ADD(v)$.
$PREV(v_c,v)$ is the closest neighbor of $v_c$ in the direction of $v$. 
}
\label{alg:struct_join}
  \STATE $v$ makes a proxy connection to node $u$.
  \STATE $v$ sends a connection request through $u$ to $ADD(v)$.
  \STATE $u$ sends a connection request to $ADD(v)$.
  \STATE $v_c$ receives the request and connects to $v$.
  \STATE $v$ sends a connection request to $PREV(v_c,v)$.
  \STATE $PREV(v_c,v)$ connects to $v$.
  \STATE $v$ is now in the correct ring location.
\end{algorithmic}
\end{algorithm}

In addition to neighbor connections, every node must also maintain $k$
shortcut
connections to other nodes that are far away in the address space.
Specifically, the distances traveled by all the shortcut connections in the
structured ring must follow a probability distribution function (pdf) of the
following form: $p(d)\propto 1/d$, where $d$ denotes the distance traveled by the
shortcut connection \cite{jk:SmallWorld,jk:Algorithmic}.
We use the local density of addresses to estimate network size
and thus $d_{ave}$, the average distance between nodes.  Then, we choose a
random distance $d$ between $d_{ave}$ and $d_{max}=2^{160}$ with probability
proportional to $1/d$
and connect to the node closest to that address using Routing
Algorithm \ref{alg:struct_greedy} (greedy routing).
The method we use to select a proper distance is to
define a random variable $x$ distributed uniformly over $[0,1]$,
and set:
\[
d = d_{ave}\left( \frac{d_{max}}{d_{ave}}\right)^x \ .
\]
From the above, we see that:
\[
Prob( d\le L) = Prob\left( x \le \frac{\log L/d_{ave}}{\log d_{max}/d_{ave}}\right)
\]
which is clearly the CDF for the random variable $d$ to be distributed
proportional to $1/d$ over $(d_{ave}, d_{max})$.
This is repeated $k$
times.
The total cost in packets to join the network is $O(\log^2 N)$, since we need
to send $O(k)$ packets and each packet requires $O(\frac{1}{k}\log^2 N)$ hops.

\section{PlanetLab Experiments}
\label{sec:experimental_results}
This section describes the results of the reliability tests of the BruNet
software.
All of the experimental results on our implementation are performed using
the global PlanetLab test-bed. PlanetLab provides a realistic, WAN environment
to test distributed applications.  In fact, PlanetLab nodes are often highly
loaded and represent a very challenging test environment.

\subsection{Experimental Methodology}
\label{sec:experimental_methodology}
PlanetLab gives access to around 400 computers
that are located in many countries around the world. There are dozens of
research projects running simultaneously on the scarce computational resources
provided by PlanetLab.  As a result, PlanetLab provides
a measure of application performance on very adverse computational and traffic load
conditions.  For the experiments presented in this section, around
100 PlanetLab machines were employed.

The current implementation is in C\# using the Mono development platform.  In order
to minimize memory and other computational resource usage on PlanetLab machines,
we run multiple nodes inside a single Mono run-time process.  As a result,
many nodes can reside on a single machine.  However,
each node is executed on a separate thread and maintains its own connections
and data.  Furthermore since class-$0$ addresses are assigned randomly, nodes
that reside on the same physical machine are unlikely to be close to each
other on the address space. We note that the UDP transport is used for all experiments presented in this
section\footnote{We have verified that the
system on a TCP transport delivers comparable performance to UDP.}.

In our experiments, we wish to see that the structure of the network is
correct, that the system can indeed route packets, and that the system is
robust to node arrivals and failures.
We analyze the logs of our experiments with a software tool which shares no
code with the BruNet system itself.
The metric we use to measure the
robustness of the network is routability.  Routability of the network
is defined as the
fraction of pairs of nodes which can communicate using the standard
(in this case greedy) routing algorithm.

\subsection{Structure Verification}
\label{sec:structure_verification}
\begin{figure}
\centering
\includegraphics[width=3in]{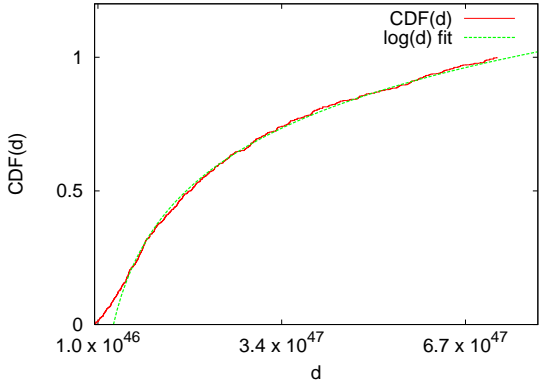}
\caption{The 1-D Kleinberg small-world structure requires that the distances of
the shortcut connections have a pdf $p(d)\propto 1/d$. In this PlanetLab
experiment, we see that the cdf(d) follows the expected logarithmic distribution
for a network of size 1060.}
\label{fig:shortcut_dist}
\end{figure}

As discussed in Sections \ref{sec:BruNet_nodes} and \ref{sec:structured_arch}, all
nodes are identified by unique 160-bit addresses, which can be interpreted
as integers;  nodes are arranged in a ring, with the convention that the integer
representation of the node addresses increase in the clockwise direction.
Furthermore, our structured small-world routing network requires that each node keeps
two neighbor connections to two closest class-$0$ addresses in the clockwise
direction and counterclockwise direction.  In other words, the structured ring
is correct if and only if the following is true: every node has
connections to its first and second class-$0$ neighbors on the clockwise and the 
counterclockwise directions in the address space.

We have successfully deployed a correct structured ring of size 1060 nodes on
PlanetLab.  It is difficult to see much in visualizations of such large graphs,
however we present several figures for various sized networks in Figure
\ref{fig:network_structure} and Figures \ref{fig:ring_join1}-\ref{fig:ring_join3}.

We verified the correctness of the shortcut distance distribution by conducting
the following: after the deployment of a correct 1060-node structured ring,
all the shortcut connection distances are extracted from the experiment
logs.  The cumulative distribution function (cdf) of the shortcut distances is plotted
in Figure \ref{fig:shortcut_dist}.  Note that the experimental cdf curve is in
good agreement with the expected curve: $cdf(d) \propto \log(d)$.

\subsection{Churn}
\label{sec:churn}
Nodes do not stay in a P2P network indefinitely.  One of the most striking
aspects of the P2P network paradigm is that we assume that nodes are fundamentally faulty
and will join and leave a network unexpectedly.  Any real system must deal
with unexpected arrivals and departures, which is called churn.

A major question is: will a node complete the joining process correctly, in
the presence of a slightly disordered network, before the node departs.  There
are two important time scales in the churn process: the mean round-trip-time
(RTT)
between the hosts at the IP layer, and the mean session time of the node.  As
the session time approaches the RTT, clearly the system will not work
properly.  Since each node requires two neighbor connections and at
least one shortcut connection, the time required to establish the node will
be much greater than the RTT.

In our experiment, we created a correct network of 980 nodes on PlanetLab.
Once the network was correct, we then started the system churning
for 25 minutes.  Each
second, with a fixed probability, every node abruptly goes offline, and
then rejoins the network.  This corresponds to an exponential distribution on
session time.

\begin{figure}
\centering
\includegraphics[width=3in]{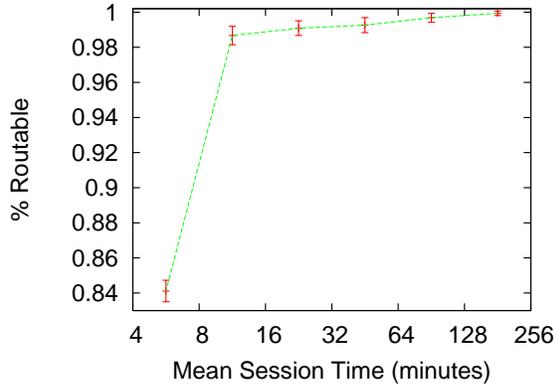}
\caption{
This experiment measures routability of a network 
of size 980 as a function the mean session time for each node.
Once mean session time is above 10 minutes, the system has nearly
perfect routability.
}
\label{fig:churn}
\end{figure}

Figure \ref{fig:churn} shows the results of our experiment.  We find that when
mean session time is above 12 minutes, the system is more than $99\%$
routable, however as mean session time decreases to 5.7 minutes, we find
that the system becomes significantly more disordered with a routability of
$84\%$.  Further decreasing the mean session time causes the system to fall
apart and tend to very low values of routability.  Exactly how the system
transitions from highly routable to non-routable is very interesting,
but is left to a future work.

Our churn model is equivalent to 
Poisson arrival and departure processes: the number of nodes
that depart in any interval is described by the Poisson distribution.
Real systems do not exhibit Poissonian churn, but instead exhibit
heavy-tailed distribution on session time: the median
uptime is often low (a few minutes) but there are many nodes
with very long uptime\cite{bamboo:usenix04}.
Simulations which have compared Poissonian churn to churn rates obtained
from real P2P traces, have found that real traces are comparable to
Poissonian churn with mean session times of around 100 minutes \cite{1009717}.
Thus, since our system can easily handle mean session times of 12 minutes,
the system should perform very well in real environments with real loads.

We note that cost of joining the network for Symphony is $O(\log^2 N)$, and
this cost comes into play when considering churn resistance.  We believe that
P2P systems with lower joining costs should
be more churn resistant.  For instance, in Viceroy\cite{dm:Viceroy}
joins cost $O(\log N)$.
Implementing Viceroy within our framework would not be difficult.

\subsection{Massive Joins and Failures}
\begin{figure*}
\centering
\includegraphics[width=6.5in]{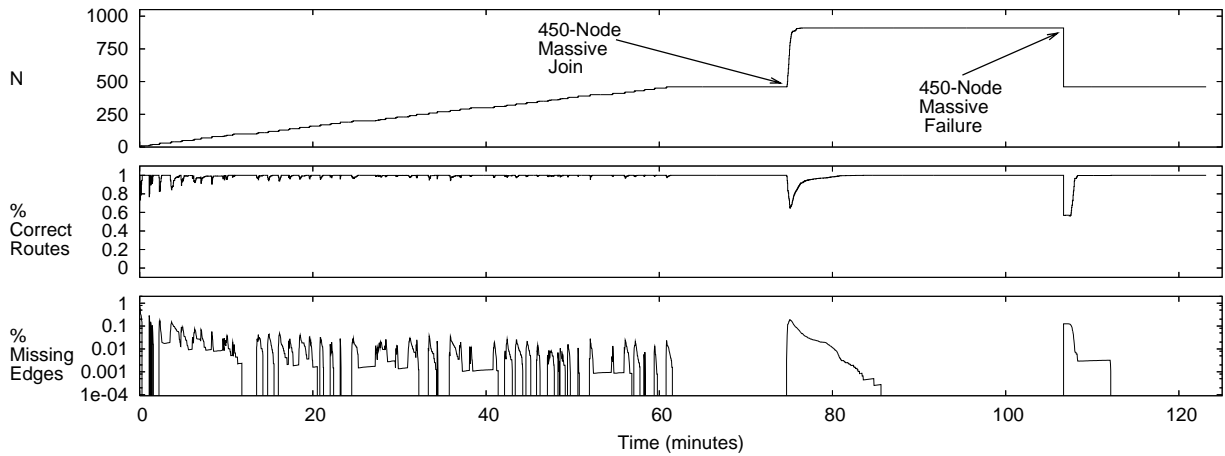}
\caption{The network is very robust during gradual joins, massive joins and massive
failures of nodes. After abrupt changes in connectivity, the network structure
heals back to a perfect ring very rapidly and achieves overwhelming percentage
routability long before the ring is completely correct. This demonstrates the
applicability of the system to highly dynamic applications.  Moreover, from examining
the bottommost figure, one can observe that the number of missing edges in the
network decreases exponentially fast in time after the massive join of 450 nodes.}
\label{fig:time_series}
\end{figure*}

One outstanding feature of this system is its ability to maintain a correct
structure under diverse node dynamics including massive node insertions,
massive node failures and even the merging of two formerly disconnected rings.
In Figure \ref{fig:time_series} we observe that nearly every pair of nodes in
the network can communicate using structured routing even under adverse 
conditions such as massive node joins and failures. 

Given that the primary objective of the presented system is overlay routing, an important
performance metric is the fraction of the pairs of nodes in the network that can communicate with 
each other; this is denoted as routability. To investigate how robust the system is to 
massive changes in network connectivity, we start with a completely routable, 460-node 
PlanetLab deployment and insert another 450 nodes into the network simultaneously. This 
experiment is depicted in Figure \ref{fig:time_series}. Less than one minute after the massive join the 
fraction of the network that is mutually routable falls to $0.65$. Within another minute 
the fraction rebounds to $0.90$. Within $11$ minutes of the massive join the entire 
910-node network is routable. 

A similar experiment was presented by Tapestry 
\cite{bz:Tapestry} where a 325-node Tapestry network experiences a 60\% massive join bringing the network size to about 525 nodes.
Prior to the massive join the routability was in the high 90\% range but not 100\% routable. 
Just after the join the routability falls below $0.70$ and then rebounds to about $0.95$ 
within $10$ minutes. However even after $60$ minutes Tapestry is still only about 95\% routable.
Thus the presented system exhibits good robustness compared to Tapestry under these
failure conditions. It should be noted that Tapestry has published fault-correcting 
protocols\cite{bz:Tapestry_Fault} designed to improve robustness under these types of
node dynamics. These additional protocols from Tapestry have been tested in a
LAN cluster but apparently not in a WAN environment such as PlanetLab.

\begin{figure*}
\centering
\begin{tabular}{c|c|c|c}
\begin{minipage}[t]{1.5in}
\includegraphics[width=1.5in]{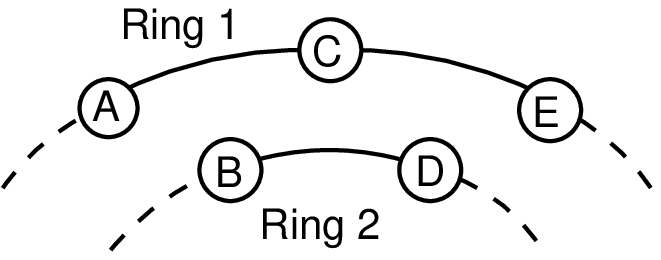}
\caption{Two distinct routable rings denoted as Ring 1 and Ring 2 can be
merged into a large routable ring.  Here we depict Ring 1 merging with Ring 2.}
\label{fig:merge_1}
\end{minipage}
& \begin{minipage}[t]{1.5in}
\includegraphics[width=1.5in]{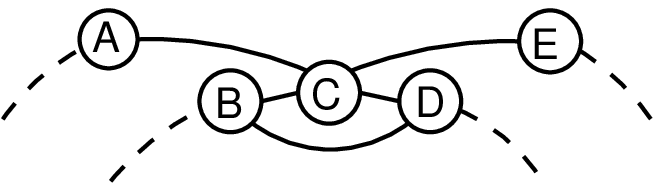}
\caption{"C" connects to "B" and "D", the two closest nodes on Ring 2. As a normal part of the connection protocol, "C" sends it neighbor lists to "B" and "D".}
\label{fig:merge_2}
\end{minipage}
& \begin{minipage}[t]{1.5in}
\includegraphics[width=1.5in]{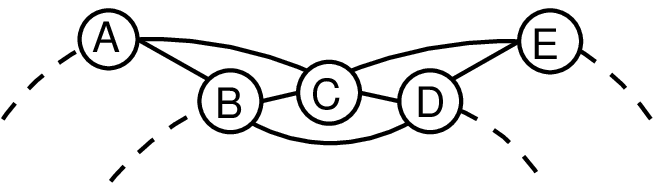}
\caption{Based on the neighbor-list information obtained from "C" while connecting, "B" connects to "A" and "D" connects to "E".}
\label{fig:merge_3}
\end{minipage}
& \begin{minipage}[t]{1.5in}
\includegraphics[width=1.5in]{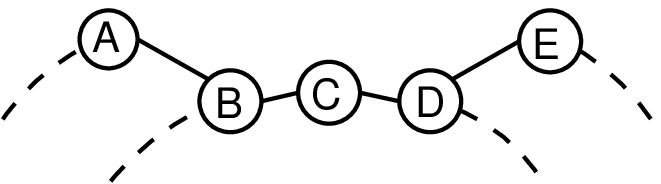}
\caption{The network is now correctly ordered but there are many more
connections than are needed. Each node maintains $k$ connections to the closest neighbors on the right and left ($k=1$ in this example). Each node will trim the excess connections until only the $k$ closest on each side remain. }
\label{fig:merge_4}
\end{minipage}\\
\end{tabular}
\end{figure*}

\begin{figure}
\begin{center}
\includegraphics[angle=90,width=2.5in]{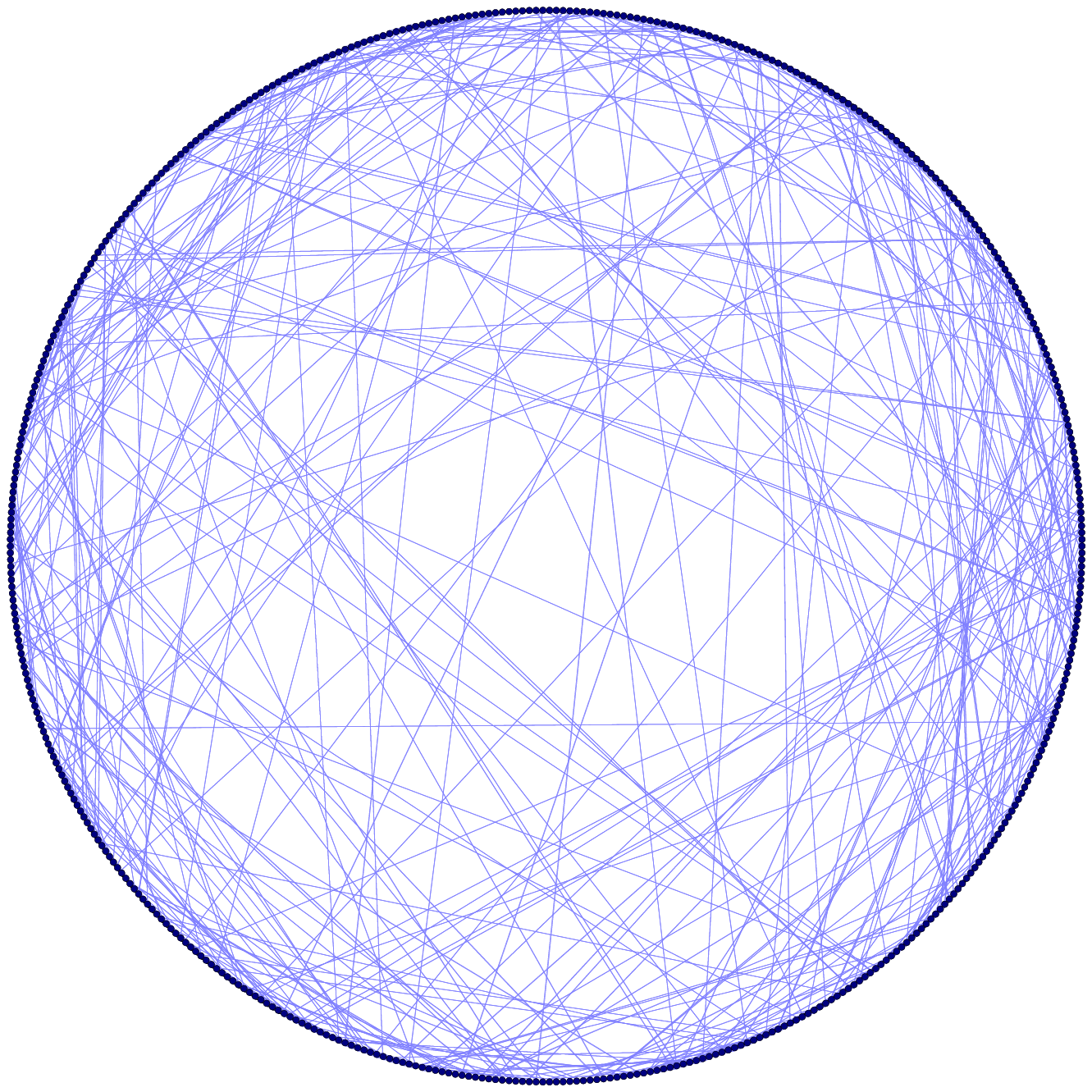}
\caption{This network on PlanetLab has $499$ nodes.}
\label{fig:ring_join1}
\end{center}
\end{figure}

\begin{figure}
\begin{center}
\includegraphics[angle=90,width=2.5in]{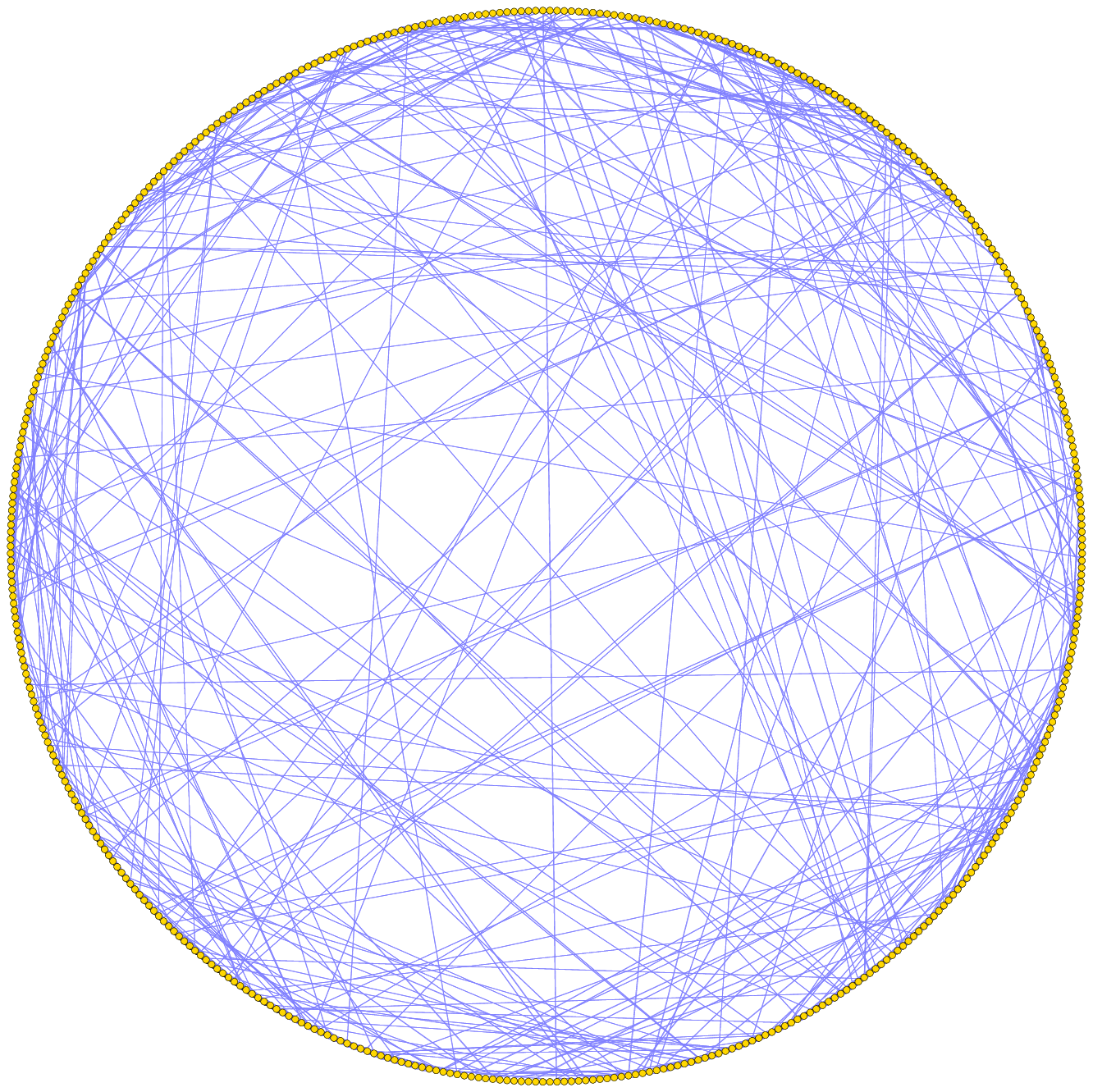}
\caption{This network on PlanetLab has $470$ nodes.}
\label{fig:ring_join2}
\end{center}
\end{figure}

\begin{figure}
\begin{center}
\includegraphics[angle=90,width=3in]{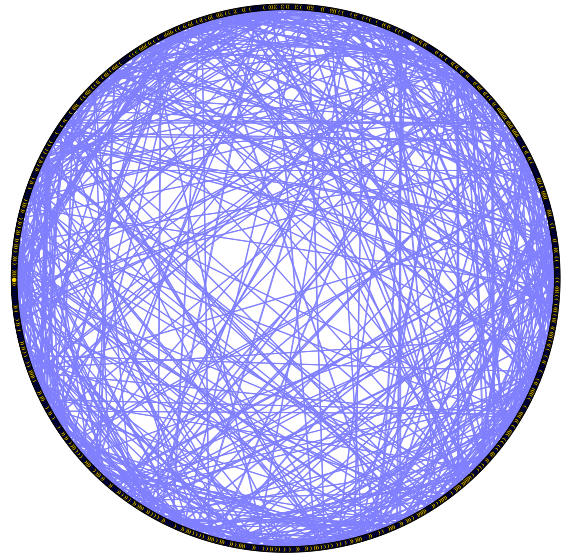}
\caption{The separate rings are merged together to form a single $970$-node
network on PlanetLab. The entire merge process takes $7$ minutes.}
\label{fig:ring_join3}
\end{center}
\end{figure}

The system can also manage the merging of multiple disconnected structured rings into a
single ring as seen in Figures \ref{fig:ring_join1}-\ref{fig:ring_join3}. 
This merging experiment was conducted as follows: we deployed two separate networks of sizes
470 and 499 respectively on PlanetLab; each network was totally unaware
of the existence of the other network (i.e. they share no nodes in common); after both
networks have formed correct rings, we deployed a single node that was
connected to nodes in both networks; as a result, the two previously disconnected rings
were merged into a single ring of size 970. The time for the two correct rings
to
merge into a single large correct ring is approximately $7$ minutes. Figures
\ref{fig:merge_1}-\ref{fig:merge_4}
show an example of how the merging dynamics works.  The exchange of neighbor
lists in the connection protocol causes the two rings to be sewn together
analogously to zipping the two halves of a zipper together. Based on this zipping
action it is clear that it will take $O(N)$ time for two rings to correctly
merge if there is a single contact point between the rings.


As demonstrated by this ring merging experiment, networks that have become
split due to catastrophic outages can easily
join back together. These findings indicate that the network will recover gracefully after
major infrastructure outages that fracture or disable large fractions of the underlying
physical layer network.

\section{Conclusion}

We present a new software framework for implementing P2P protocols.
We use this framework to present the first 1-D implementation of the Kleinberg
routable small-world model.  We have shown that the
C\# implementation produces networks that have the required topological structure
To provide scalable structured small-world routing. The system is also very robust
in the presence of large node dynamics including
massive joins, massive failures, disconnected ring merges and churn. Given
that this system is intended to provide overlay routing over heterogeneous
physical layers and transport protocols, this robustness is critical to
enabling reliable overlay applications.

We anticipate that this framework
will be valuable to other researchers to allow them to implement new P2P
routing and connection management protocols, without the need to reimplement
solutions to common problems of node handshaking, packet sending and
receiving, and abstraction of underlying transports, such as UDP and TCP.
Future work will including using this framework to implement unstructured
P2P protocols along with structured P2P protocols.

\section{Acknowledgements}
We would like to thank Nikolas Kontorinis and Julie Walters for many helpful comments.

\bibliographystyle{IEEEtran}
\bibliography{brunet-paper0,bib_brunet}

\end{document}